# A New Capacity Result for the Z-Gaussian Cognitive Interference Channel

Stefano Rini, Daniela Tuninetti and Natasha Devroye
ECE Deptartment, University of Illinois at Chicago, Chicago, IL 60607, USA,
Email: srini2, danielat, devroye@uic.edu

*Abstract*—This work proposes a novel outer bound for the Gaussian cognitive interference channel in strong interference at the primary receiver based on the capacity of a multi-antenna *broadcast channel with degraded message set*. It then shows that for the Z-channel, i.e., when the secondary receiver experiences no interference and the primary receiver experiences strong interference, the proposed outer bound not only is the tightest among known bounds but is actually achievable for sufficiently strong interference. The latter is a novel capacity result that from numerical evaluations appears to be generalizable to a larger (i.e., non-Z) class of Gaussian channels.

## I. INTRODUCTION

Cognitive radio is a novel paradigm for wireless networks whereby a node changes its communication scheme to efficiently share the spectrum with licensed and unlicensed users. The actual communication scheme used depends on the overall instantaneous network activity, which the cognitive device is assumed to be able to track. In its simplest form a cognitive network is modeled as a two-user interference channel, where one transmitter-receiver pair is referred to as the *primary* pair and the other as the *cognitive/secondary* pair. The primary transmitter has knowledge of one of the two independent messages to be sent, while the cognitive transmitter has full, non-causal knowledge of both messages, thus idealizing the cognitive user's ability to detect network activity.

### A. Past work

The information theoretic capacity of the cognitive interference channel (CIFC) remains elusive in general. The CIFC was first considered in [1], where an achievable rate region (for general discrete memoryless channels) and a broadcast-channel outer bound (in Gaussian noise only) were proposed.

**Inner bounds:** Since the cognitive transmitter can "broadcast" information to both receivers, achievable strategies for the CIFC contains features of both the interference and of the broadcast channel, such as rate splitting, superposition coding, binning and simultaneous decoding. A comparison of all the transmission schemes proposed in the literature was presented in [2], in which we showed that our region in [2, Th.5.1] is provably the largest known achievable rate region to date.

**Outer bounds:** The tightest known outer bound for the general CIFC was derived in [3, Th.4] using a technique originally developed for the broadcast channel in [4]. Both the "weak interference" outer bound of [5, Th.3.2] and the "strong interference" outer bound of [6, Th.4] may be derived by loosening [3, Th.4]. The outer bound in [3, Th.4] is however difficult to evaluate because it contains three auxiliary random variables for which no cardinality bounds are given on the corresponding alphabets. Moreover, for the Gaussian channel, the "Gaussian maximizes entropy" property alone does not suffice to show that Gaussian inputs exhaust the outer bound. For these reasons, in [2, Th.4.1] we proposed an outer bound that exploits the fact that the capacity region only depends on the conditional marginal distributions (as for broadcast channels [7], since the receivers do not cooperate). The resulting outer bound does not include auxiliary random variables and every mutual information term involves all the inputs, like in the cut-set bound [8, Th.15.10.1]; this implies that it may be easily evaluated for many channel of interest, including the Gaussian channel.

**Capacity Results:** The first capacity results for the CIFC were determined in [5, Th.3.4] for channels with "very weak interference" at the primary receiver and in [9, Th.6] for channels with "very strong interference". In [2, Th.7.1], we showed that the outer bound of [5, Th.3.2] is achievable in what we termed the "better cognitive decoding" regime, which includes both the "very weak interference" and the "very strong interference" regimes and is the largest class of discrete memoryless CIFCs for which capacity is known.

For the Gaussian CIFC (G-CIFC), capacity in "weak interference" was determined in [5, Th.3.7] and in independently in [10, Th.4.1], and in "very strong interference" in [9, Th.6]. In [2, Th.4.1] we proposed a unified derivation of the outer bounds for the "weak interference" and for the "strong interference" regimes of [5, Th.3.7] and [3, Th.5], respectively. Moreover, in [11, Th.3.1] we showed that the outer bound in [2, Th.4.1] is achievable also in the *primary decodes cognitive* regime, which only in part coincides with the "very strong interference" regime for which capacity was known [6, Th.6]. The outer bound in [2, Th.4.1] is also tight for the class of *semi-deterministic CIFCs with a noiseless output at the primary receiver* [12, Th.2], and is capacity to within 0.5 bit/s/Hz per real-dimension for any G-CIGC [12, Th.3] (thus improving on our previous constant gap result of 1.87 bit/s/Hz per real-dimension in [13, Sec. IV]).

**Z-channel:** The special case where only one receiver experiences interference is known as the Z-channel. For the case where the cognitive transmitter does not create interference to the primary receiver and the cognitive-primary link is noiseless, inner and outer bounds were obtained in [14]; the Gaussian counterpart is trivial. For the case where the primary transmitter does not create interference to the secondary receiver capacity is known by specializing the "weak interference" result of [5, Th.3.7] or the "primary decodes cognitive" result of [11], [15, Th.3.1]; capacity remains open for sufficiently strong interference.

### B. Contributions and Paper Organization

This paper presents two main results:

1) We first propose a novel outer bound for the G-CIFC with strong interference at the primary receiver based on enhancing the original channel into a *multi-antenna broadcast channel with degraded message set*. By using the "extremal inequality" of [16], we show that Gaussian input is optimal for the novel bound.
2) For the Z-G-CIFC with strong interference (where the secondary receiver does not experience interference and the primary receiver experiences strong interference) we show that there exists a set of parameters where our novel outer bound is the tightest known. We then propose an achievable scheme based on superposition coding that meets the novel outer bound for sufficiently strong interference thus proving a new capacity result.

The rest of the paper is organized as follows. Section II defines the G-CIFC and reports the outer bound of [11, Th.2.2]. In Section III we present our novel outer bound and in Section IV we prove a new capacity result the Z-G-CIFC with strong interference. We also show in Section V by means of a numerical example that the proposed outer bound meets an achievable scheme – up to Matlab numerical precision – for some general (i.e., non-Z) G-CIFC; for these channels however a formal proof of capacity is not yet available. Section VI concludes the paper. The Appendix contains some of the proofs. Our notation follows the convention of [17].

## II. CHANNEL MODEL

### A. Gaussian Channel (G-CIFC)

A two-user complex-valued G-CIFC in *canonical form* [2], as depicted in Fig. 1, has outputs:

$$Y_1 = X_1 + aX_2 + Z_1,$$
$$Y_2 = |b|X_1 + X_2 + Z_2,$$

where the channel gains $a$ and $b$ are constant and known to all terminals, the inputs are subject to the power constraint:

$$\mathbb{E}[|X_i|^2] \leq P_i, \quad P_i \in \mathbb{R}^+, \quad i \in \{1, 2\},$$

and the noise $Z_i$ is $\mathcal{N}(0,1)$, $i \in \{1, 2\}$. Each transmitter $i$, $i \in \{1, 2\}$, wishes to communicate a message $W_i$, uniformly distributed on $[1 : 2^{NR_i}]$, to receiver $i$ in $N$ channel uses at rate $R_i$. The two messages are independent. Message $W_2$ is also available to transmitter 1. A rate pair $(R_1, R_2)$ is achievable if there exists a sequence of encoding functions $X_1^N(W_1, W_2)$ and $X_2^N(W_2)$, and a sequence of decoding functions $\widehat{W}_i(Y_i^N)$, $i \in \{1, 2\}$, such that the error probability vanishes as $N \to \infty$. The capacity region is the convex closure of the set of achievable rates [8] and is indicated with $\mathbf{C}(a, |b|, P_1, P_2)$. The capacity $\mathbf{C}(a, |b|, P_1, P_2)$ is not known in general.

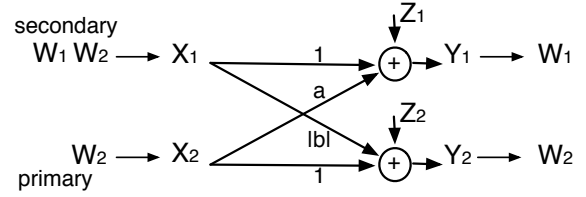

Fig. 1. The general Gaussian Cognitive Interference Channel (G-CIFC).

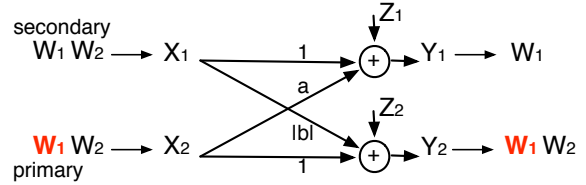

Fig. 2. The Broadcast Channel with Degraded Message Set (BC-DMS) that outer bounds the G-CIFC in Fig. 1 for $|b| \geq 1$.

### B. Gaussian Z-Channel (Z-G-CIFC)

A G-CIFC is said to be a Z-channel if either $a$ or $b$ are zero. If $|b| = 0$, i.e., the primary receiver does not experience interference from the cognitive transmitter, the capacity is trivially given by

$$\mathbf{C}(a, 0, P_1, P_2) = \{R_1 \leq \log(1 + P_1), \ R_2 \leq \log(1 + P_2)\}.$$

If $a = 0$, i.e., the cognitive receiver does not experience interference from the primary transmitter, the capacity is only known for $|b| \leq 1$ [5] and for $1 < |b| \leq \sqrt{1 + \frac{P_2}{P_1+1}}$ [11]; in both cases $\mathbf{C}(0, |b|, P_1, P_2)$ is given in (1) below. The case $|b| > \sqrt{1 + \frac{P_2}{P_1+1}}$ will be used in Section IV as a case-study for our novel outer bound developed in Section III.

### C. Known Outer Bound

The best known *computable* outer bound for the G-CIFC is given by the "unifying outer bound" of [11, Th.2.2], i.e.,

$$R_1 \leq \log(1 + \alpha P_1), \tag{1a}$$
$$R_2 \leq \log\left(1 + |b|^2 P_1 + P_2 + 2\sqrt{\bar{\alpha}|b|^2 P_1 P_2}\right), \tag{1b}$$
$$R_1 + R_2 \leq \log\left(1 + |b|^2 \bar{\alpha} P_1 + P_2 + 2\sqrt{\bar{\alpha}|b|^2 P_1 P_2}\right)$$
$$+ [\log(1 + \alpha P_1) - \log(1 + |b|^2 \alpha P_1)]^+ \tag{1c}$$

taken over the union of all $\alpha \in [0, 1]$.

**Remark:** When is the outer bound in (1) tight? In "strong interference" ($|b| > 1$) the region in (1) reduces to the outer bound [3, Th.4], which is tight in "very strong interference" [9, Th.6]. The bound in (1) is also tight in "weak interference" ($|b| \leq 1$) [5, Lemma 3.6] and [10, Th.4.1] as well as in the "primary decodes cognitive" regime [11, Th.3.1]. For other parameter values, the bound in (1) is capacity to within

0.5 bits/s/Hz per real-dimension [12, Th.3] and to within a factor two [12, Th.4].

**Remark:** Can the outer bound in (1) be tight in general? The bound in (1) is not tight in general. To see this, consider the case $|b| > 1$ and $P_2 = 0$ (the primary user is silent). This channel is equivalent to a degraded broadcast channel (BC) with input $X_1$ whose capacity $\mathbf{C}(a, |b|, P_1, 0)$ given by [18]:

$$R_1 \leq \log\left(1 + \frac{\alpha P_1}{\bar{\alpha} P_1 + 1}\right), \ R_2 \leq \log\left(1 + |b|^2 \bar{\alpha} P_1\right),$$

for all $\alpha \in [0, 1]$, with $\bar{\alpha} = 1 - \alpha$. For $P_2 = 0$ and $|b| > 1$ the outer bound in (1) reduces to:

$$R_1 \leq \log(1 + P_1), \ R_1 + R_2 \leq \log(1 + |b|^2 P_1).$$

It is easy to see that the latter region fully contains the former. The derivation of a bound that is tighter than (1) in strong interference ($|b| > 1$) is the first goal this paper.

## III. BC-BASED OUTER BOUND

In this section we propose an outer bound that is tighter than (1). The following observation is key: if we provide the primary transmitter with the cognitive message, the CIFC becomes a BC with input $X = (X_1, X_2)$; thus, an outer bound valid for a general (not necessarily Gaussian) CIFC is:

$$\mathcal{R}^{(\text{BC-PR})} \cap \mathcal{R}^{(\text{CIFC})}, \qquad (2)$$

where $\mathcal{R}^{(\text{BC-PR})}$ is the capacity region (or an outer bound) for the *BC with private rates only* and where $\mathcal{R}^{(\text{CIFC})}$ is any outer bound for the CIFC. For the G-CIFC the bound in (2) is as follows: $\mathcal{R}^{(\text{BC-PR})}$ is the capacity with Private Rates (PR) only of MIMO BC with two antennas at the transmitter, one antenna at each receiver, and with a per-antenna power constraint, as originally used in [1, page 1819], and $\mathcal{R}^{(\text{CIFC})}$ is given in (1).

The bound in (2) may be further tightened for the G-CIFC in the "strong interference" ($|b| > 1$) regime as follows. As previously noted in [3, Sec. 6.1], in the "strong interference" regime there is no loss of optimality in having the primary receiver decode the cognitive message in addition to its own message. Indeed, after decoding $W_2$, receiver 2 can reconstruct $X_2^N(W_2)$ and compute the following estimate of $Y_1$

$$\widetilde{Y}_1^N \triangleq \frac{Y_2^N - X_2^N}{|b|} + aX_2^N + \sqrt{1 - \frac{1}{|b|^2}} Z_0^N \sim Y_1^N,$$

where $Z_0^N \sim \mathcal{N}(0, \mathbf{I})$ and independent of everything else. Hence, if the secondary receiver can decode $W_1$ from $Y_1^N$, so can the primary receiver from $\widetilde{Y}_1^N$. For this reason the capacity region of the G-CIFC with $|b| > 1$ is unchanged if receiver 2 is required to decoded both messages. If we further allow the two transmitters to fully cooperate, the resulting channel is a *Gaussian MIMO BC with Degraded Message Set (DMS)* (see Fig. 2), where message $W_2$ is to be decoded at receiver 2 only and message $W_1$ at both receivers. The capacity of the Gaussian MIMO BC-DMS and an input covariance constraint was determined in [19, Th.5]. Following from the previous discussion:

**Theorem 1. BC-DMS-based outer bound for the G-CIFC.** *The capacity of a G-CIFC in "strong interference" ($|b| > 1$) satisfies:*

$$\mathbf{C}(a, |b|, P_1, P_2) \subseteq \mathcal{R}^{(\text{BC-DMS})} \cap \mathcal{R}^{(\text{SI})}, \qquad (3)$$

*where $\mathcal{R}^{(\text{BC-DMS})}$ is the capacity of the Gaussian MIMO BC with degraded message set and $\mathcal{R}^{(\text{SI})}$ is the outer bound in (1) for $|b| > 1$.*

The analytical evaluation of the outer bound in (3) for a general G-CIFC is quite involved. For the special case of Z-G-CIFC (i.e., $a = 0$) a closed form expression may be obtained as follows (the proof may be found in the Appendix):

**Corollary 2. BC-DMS-based outer bound for the Z-G-CIFC.** *For a G-CIFC with $a = 0$ and $|b| \geq 1$ the outer bound in (3) is contained into the region:*

$$R_1 \leq \log(1 + \alpha P_1), \qquad (4a)$$

$$R_2 \leq \log\left(1 + \left(\sqrt{P_2} + \sqrt{\frac{|b|^2 P_1 \bar{\alpha}}{1 + \alpha P_1}}\right)^2\right), \qquad (4b)$$

$$R_1 + R_2 \leq \log\left(1 + P_2 + |b|^2 P_1 + 2\sqrt{\bar{\alpha}|b|^2 P_1 P_2}\right). \qquad (4c)$$

*Moreover, the $R_2$-bound from the MIMO BC-DMS outer bound (from (4b)) is more stringent than the $R_2$-bound from the "strong interference" outer bound (from the difference of (4c) and (4a)) if*

$$|b| \geq \sqrt{P_2 + 1}. \qquad (5)$$

## IV. NEW CAPACITY RESULT

By using the outer bound of Corollary 2, together with the general achievable region of [2, Sec. VIII], we have:

**Theorem 3. Capacity for some Z-G-CIFCs.** *For a G-CIFC with $a = 0$ and with*

$$|b| \geq \sqrt{1 + P_2(1 + P_1)} + \sqrt{P_1 P_2} \qquad (6)$$

*the outer bound in Corollary 2 is tight.*

*Proof:* We consider a simple superposition coding scheme [2, Scheme (E), Sec.VIII]. Encoding: let $X_2 \sim \mathcal{N}(0, P_2)$ and $X_1 = \sqrt{(1-\beta)P_1/P_2} X_2 + \sqrt{\beta P_1} U_{1c}$, with $U_{1c} \sim \mathcal{N}(0, 1)$ independent of $X_2$, and with $\beta \in [0, 1]$. Decoding: decoder 2 jointly decodes $X_2$ and $U_{1c}$ from

$$Y_2 = \left(1 + \sqrt{(1-\beta)|b|^2 P_1/P_2}\right) X_2 + \sqrt{\beta|b|^2 P_1} U_{1c} + Z_2;$$

decoder 1 only decodes $U_{1c}$ by treating $X_2$ as noise from

$$Y_1 = \sqrt{(1-\beta)P_1/P_2} X_2 + \sqrt{\beta P_1} U_{1c} + Z_2.$$

The achievable region is:

$$R_1 \le \log\left(1 + \frac{\beta P_1}{1+(1-\beta)P_1}\right), \quad (7a)$$
$$R_2 \le \log\left(1 + (\sqrt{P_2} + \sqrt{(1-\beta)|b|^2 P_1})^2\right), \quad (7b)$$
$$R_1 + R_2 \le \log\left(1 + P_2 + |b|^2 P_1 + 2\sqrt{(1-\beta)|b|^2 P_1 P_2}\right). \quad (7c)$$

Let now $\frac{\beta}{1+(1-\beta)P_1} = \alpha$, that is, $\beta = \frac{1+P_1}{1+\alpha P_1}$; with this choice we have (7a)=(4a) and (7b)=(4b). If we show that when the sum-rate in (7c) is redundant when the condition in (5) is met then we have shown that the achievable region in (7) coincides with the outer bound in (4); this is the case if

$$1 + P_2 + |b|^2 P_1 + 2\sqrt{(1-\beta)|b|^2 P_1 P_2}$$
$$\ge \frac{1+P_1}{1+(1-\beta)P_1}(1 + P_2 + |b|^2 P_1 - \beta|b|^2 P_1$$
$$+ 2\sqrt{(1-\beta)|b|^2 P_1 P_2}), \forall \beta \in [0,1]$$
$$\iff |b|^2 \ge 1 + P_2 + 2\sqrt{(1-\beta)|b|^2 P_1 P_2}, \forall \beta \in [0,1],$$
$$\iff |b|^2 \ge 1 + P_2 + 2\sqrt{|b|^2 P_1 P_2},$$

which corresponds to (6). Clearly the regime identified by (6) is such that the condition is (5) is met. QED. ∎

**Remark:** Our capacity result in Th.3 and our previous capacity result in [11, Th.3.1] imply that the capacity of the Z-G-CIFC with

$$|b| \in \left[\sqrt{1 + P_2\left(1 - \frac{P_1}{P_1+1}\right)}, \sqrt{1+P_2(1+P_1)} + \sqrt{P_1 P_2}\right]$$

is still open, i.e, in this regime the capacity is only known to within 0.5 bit/s/Hz per real-dimension [12, Th.3].

## V. NUMERICAL RESULT

For a general G-CIFC (with $a \ne 0$) it is challenging to analytically show that the outer bound region in Th.1 meets the general inner bound region in [2, Sec. IV] due to the numerous parameters involved in determining the points on the convex closure of the inner and outer bounds. In Fig. 3 we show the result of the numerical optimization of the outer bound region in Th.1 and of the general inner bound region in [2, Sec. IV] for $a = 0.01$, $|b| = 10$, $P_1 = P_2 = 5$. We see that the inner and outer bounds coincides up to Matlab numerical precision. Although this does not constitute a formal proof of capacity, it shows that our outer bound region in Th.1 could be capacity for a more general class of C-IFC than that identified by Th.3.

## VI. CONCLUSION

In this paper we proposed a novel outer bound for the Gaussian cognitive interference channel in strong interference by noticing that in this regime the channel may be enhanced to a MIMO BC with degraded message set if the transmitters are allowed to cooperate. For the special case of the Z-Gaussian cognitive interference channel we showed that the proposed bound is tighter than existing ones for certain parameter regimes and that it is capacity for sufficiently strong interference.

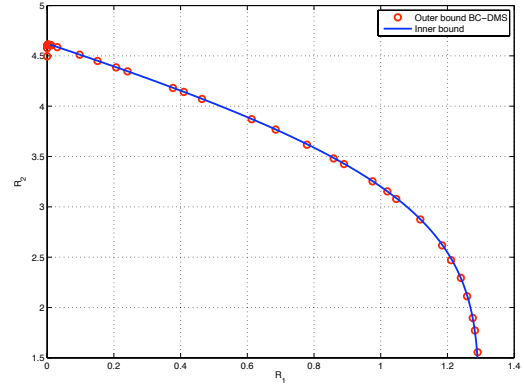

Fig. 3. The outer bound in Th.1 and the general achievable region of [2, Sec. IV] for the G-CIFC with $a = 0.01$, $|b| = 10$, $P_1 = P_2 = 5$. Notice the the vertical axis starts at 1.5.

## APPENDIX

The capacity region of the general BC-DMS where receiver 2 must decode both messages is [20]

$$R_1 \le I(U; Y_1), \quad (8a)$$
$$R_2 \le I(X; Y_2|U), \quad (8b)$$
$$R_1 + R_2 \le I(X; Y_2), \quad (8c)$$

for all joint distributions $P_{U,X}$. A closed form expression for (8) for the Gaussian MIMO BC-DMS was derived in [21]; it was however not obtained as a direct computation of (8) but was instead expressed as the intersection of the capacity region of a general Gaussian MIMO BC-PR and an additional sum-rate constraint; the evaluation of the Gaussian MIMO BC-PR region involves maximization over covariance matrices and "dirty paper coding" orders and is thus very difficult to carry out in closed form. By using the recent "extremal inequality" result of [16, Th.1], together with a series of steps as in [19], it is not difficult to show that jointly Gaussian $(U, X)$ are optimal in (8) for a general Gaussian MIMO BC-DMS with an arbitrary input covariance constraint; this result, formally proved in [22, Th.3.3], greatly simplifies the evaluation of (8).

We now evaluate the region in (8) for the following jointly Gaussian input. For an input covariance $\text{Cov}[X] = \mathbf{S}$, with:

$$\mathbf{S} \triangleq \begin{pmatrix} P_1 & \rho\sqrt{P_1 P_2} \\ \rho^*\sqrt{P_1 P_2} & P_2 \end{pmatrix} \quad (9)$$

let $U \sim \mathcal{N}(0, \mathbf{B}_1)$ independent of $V \sim \mathcal{N}(0, \mathbf{B}_2)$, and let $X = U + V$, with:

$$\mathbf{B}_1 = \begin{bmatrix} \alpha_1 P_1 & \rho_1\sqrt{\alpha_1 P_1\, \alpha_2 P_2} \\ \rho_1^*\sqrt{\alpha_1 P_1\, \alpha_2 P_2} & \alpha_2 P_2 \end{bmatrix},$$
$$\mathbf{B}_2 = \begin{bmatrix} \bar{\alpha}_1 P_1 & \rho_2\sqrt{\bar{\alpha}_1 P_1\, \bar{\alpha}_2 P_2} \\ \rho_2^*\sqrt{\bar{\alpha}_1 P_1\, \bar{\alpha}_2 P_2} & \bar{\alpha}_2 P_2 \end{bmatrix},$$

such that:

$$(\alpha_1, \alpha_2, |\rho_1|, |\rho_2|) \in [0,1]^4: \ \rho_1\sqrt{\alpha_1 \alpha_2} + \rho_2\sqrt{\bar{\alpha}_1 \bar{\alpha}_2} = \rho.$$

The condition $(\alpha_1, \alpha_2) \in [0,1]^2$ is to guarantee that the per-antenna power constraint is met.

Since (8a) and (8b) correspond to the DPC (dirty paper coding) region for a BC-PR (with user 2 encoded last) and since the sum-rate in (8c) depends only on the parameter $\rho$ in (9), we write the BC-DMS region for the equivalent BC with channel matrices $\mathbf{h}_1 = [1\ 0]$ and $\mathbf{h}_2 = [|b|\ 1]$ as:

$$\mathcal{R}^{(\text{BC-DMS})} = \bigcup_{|\rho|\leq 1\ \alpha_1\in[0,1]} \left(\mathcal{R}^{(\text{DPC})}(\rho,\alpha_1) \cap \mathcal{R}^{(\text{sum})}(\rho)\right)$$

$$\subseteq \bigcup_{|\rho|\leq 1, \alpha_1\in[0,1]} \left\{ \begin{array}{l} R_1 \leq R_1^{(\text{DPC})}(\alpha_1) \\ R_2 \leq R_2^{(\text{DPC})}(\rho,\alpha_1) \end{array} \right.$$

where the region $\mathcal{R}^{(\text{BC-DMS})}$ has been enlarged by removing

$$\mathcal{R}^{(\text{sum})}(\rho)$$
$$= \left\{ R_1 + R_2 \leq \log(1 + |b|^2 P_1 + P_2 + 2\text{Re}\{\rho\}\sqrt{|b|^2 P_1 P_2}) \right\}$$

and where the region $\mathcal{R}^{(\text{DPC})}(\rho,\alpha_1)$ is defined by:

$$R_1^{(\text{DPC})}(\alpha_1) \triangleq \log\left(1 + \frac{\alpha_1 P_1}{1 + \bar{\alpha}_1 P_1}\right), \tag{10a}$$

$$R_2^{(\text{DPC})}(\rho,\alpha_1) \triangleq \max_{\rho_1,\rho_2,\alpha_2\ \text{s.t.}\ \rho=\rho_1\sqrt{\alpha_1\alpha_2}+\rho_2\sqrt{\bar{\alpha}_1\bar{\alpha}_2}}$$
$$\log(1 + |b|^2 \bar{\alpha}_1 P_1 + \bar{\alpha}_2 P_2 + 2\text{Re}\{\rho_2\}\sqrt{\bar{\alpha}_1\bar{\alpha}_2|b|^2 P_1 P_2})$$
$$\leq \log(1 + |b|^2 \bar{\alpha}_1 P_1 + P_2 + 2\sqrt{\bar{\alpha}_1|b|^2 P_1 P_2}), \tag{10b}$$

where the inequality follows by optimizing over $(\alpha_2, |\rho_1|, |\rho_2|) \in [0,1]^3$ without accounting for the constraint.

The inequalities in (10) prove that for the Z-G-CIFC, $\mathcal{R}^{(\text{DPC})}(\rho,\alpha_1)$, and thus also $\mathcal{R}^{(\text{BC-DMS})}$, is contained in the region:

$$R_1 \leq \log(1 + \alpha P_1) \triangleq R_1^{(\text{BC-DMS-Z})}, \tag{11a}$$

$$R_2 \leq \log\left(1 + \left(\sqrt{|b|^2 \frac{(1-\alpha)P_1}{1+\alpha P_1}} + \sqrt{P_2}\right)^2\right) \triangleq R_2^{(\text{BC-DMS-Z})} \tag{11b}$$

taken over the union of all $\alpha \in [0,1]$, with the "change of variable": $\alpha = \frac{\alpha_1}{1+\bar{\alpha}_1 P_1}$.

Finally, the BC-DMS outer bound of (11) is more stringent than the outer bound in (1) if

$$R_1^{(\text{BC-DMS-Z})} + R_2^{(\text{BC-DMS-Z})} \leq R_{\text{sum}}^{(\text{SI})}\ \forall\alpha \in [0,1]$$
$$\iff \alpha P_1 + \left(\sqrt{|b|^2(1-\alpha)P_1} + \sqrt{P_2(1+\alpha P_1)}\right)^2$$
$$\leq P_2 + |b|^2 P_1 + 2\sqrt{(1-\alpha)|b|^2 P_1 P_2}\ \ \forall\alpha \in [0,1]$$
$$\iff 1 + P_2 - |b|^2 \leq 0,$$

as claimed in (5). QED.

### ACKNOWLEDGMENT

The work of the authors was partially funded by NSF under awards number 0643954 and 1017436. The contents of this article are solely the responsibility of the authors and do not necessarily represent the official views of the NSF. The authors would like to acknowledge many helpful discussions with Prof. Shlomo Shamai.